\documentstyle[aps,epsf]{revtex}

\begin{document}

\title{Maximum Mass-Radius Ratios for Charged Compact General Relativistic Objects}
\author{M.K. Mak\footnote{E-mail:mkmak@vtc.edu.hk} and Peter N. Dobson, Jr.\footnote{E-mail:aadobson@ust.hk}}    
\address{Department of Physics, The University of Science and Technology,\\
Clear Water Bay, Hong Kong, P. R. China}
\author{T. Harko\footnote{E-mail:tcharko@hkusua.hku.hk}}
\address{Department of Physics, The University of Hong Kong, Pokfulam Road, Hong Kong, P. R. China}

\maketitle

\begin{abstract}
Upper limits for the mass-radius ratio and total charge are derived for
stable charged general relativistic matter distributions.
For charged compact objects the mass-radius ratio exceeds the value
4/9 corresponding to neutral stars.
General restrictions for the red shift and total energy
(including the gravitational contribution) are also
obtained.

\end{abstract}

Pacs Numbers: 04.20.-q, 97.60.-s

% \date{July 17, 2000}

\section{Introduction}

It is generally accepted today that black holes are uniquely characterized
by their total mass-energy $E$, charge $Q$ and angular momentum $J$ \cite{Ch99}. Most of the investigations of the astrophysical objects have
been done under the assumption of the electric charge neutrality of the
stellar matter. However, as a result of accretion
neutron stars can acquire a net charge, if accretion produces luminosity
close to the Eddington limit $L_{E}$  \cite{Sh71}. Let us consider a
star of mass $M$ that undergoes spherical accretion and assume, for
simplicity, that the accreting material is ionized hydrogen. If the
accreting luminosity is $L$, the infalling electrons experience a radiative
force $F_{R}=\frac{\sigma _{T}L}{4\pi cr^{2}}$, where $\sigma _{T}$ is the
Thomson cross section. Since the radiation drag acting on the protons is a
factor $\left( m_{e}/m_{p}\right) ^{2}$ smaller, electrons and protons are
subject to different accelerations, and the star acquires a net positive
charge $Q=\frac{GMm_{p}}{e}\frac{L}{L_{E}}$  \cite{Tu97}, where $%
L_{E}=4\pi GMm_{p}c/\sigma _{T}$ is the Eddington luminosity. The
astrophysical conditions under which this phenomenon can take place are
rather extreme but in principle they could lead to a charged astrophysical
configuration. This mechanism has been recently proposed, via vacuum
breakdown near a charged black hole, as a source of $\gamma $-ray bursts
\cite{Pr98}. A phase transition of neutron matter to quark matter at
zero temperature or temperatures small compared to degeneracy temperature
allows the existence of hybrid stars, i.e. stars having a quark core and a
crust of neutron matter \cite{Wi84}. In fact, quark matter with
electrically charged constituents rather than neutron matter could
hold the large magnetic field of the pulsars \cite{Kh95} and hence it
is possible that for strange-matter made stars the effects of the non-zero
electrical charge be important.

By using the static spherically symmetric gravitational field equations
Buchdahl \cite{Bu59} has obtained an absolute constraint of the
maximally allowable mass $M$- radius $R$ for isotropic fluid
spheres of the form $\frac{2M}{R}<\frac{8}{9}$ (where natural units $c=G=1$
have been used).

It is the purpose of the present Letter to obtain the maximum allowable mass
-radius ratio in the case of stable charged compact general relativistic objects.
This is achieved by generalizing to the charged case the method described for neutral stars in
Buchdahl \cite{Bu59} and Straumann \cite{St84}.

\section{Maximum mass-radius ratio for charged general relativistic compact objects}

For a static general relativistic spherically symmetric configuration the
interior line element is given by $ds^{2}=e^{\nu }dt^{2}-e^{\lambda }dr^{2}-r^{2}\left( d\theta ^{2}+\sin
^{2}\theta d\varphi ^{2}\right) $.

The properties of the charged compact object can be completely described by
the structure equations, which are given by \cite{Be71}:
\begin{equation}\label{1}
\frac{dm}{dr}=4\pi \rho r^{2}+\frac{Q}{r}\frac{dQ}{dr},
\end{equation}
\begin{equation}\label{2}
\frac{dp}{dr}=-\frac{\left( \rho +p\right) \left( m+4\pi r^{3}p-\frac{%
Q^{2}}{r}\right) }{r^{2}\left( 1-\frac{2m}{r}+\frac{Q^{2}}{r^{2}}\right) }+%
\frac{Q}{4\pi r^{4}}\frac{dQ}{dr},
\end{equation}
\begin{equation}\label{3}
\frac{d\nu }{dr}=\frac{2\left( m+4\pi r^{3}p-\frac{Q^{2}}{r}\right) }{%
r^{2}\left( 1-\frac{2m}{r}+\frac{Q^{2}}{r^{2}}\right) },
\end{equation}
where $\rho $ is the energy density of the matter, $p$ is the thermodynamic
pressure, $m(r)$ is the mass and $Q(r)=4\pi \int_{0}^{r}e^{\frac{\nu +\lambda }{2}%
}r^{2}j^{0}dr$ is the charge inside radius $r$, respectively. The
electric current inside the star is given by $j^{i}=\left(
j^{0},0,0,0\right) $ .

The structure equations (\ref{1})-(\ref{3}) must be considered together with
the boundary conditions $p(R)=0$, $p(0)=p_{c}$ and $\rho (0)=\rho _{c}$,
where $\rho _{c}$, $p_{c}$ are the central density and pressure, respectively.

With the use of Eqs. (\ref{1})-(\ref{3}) it is easy to show that the function
$\zeta =e^{\frac{\nu }{2}}>0,\forall r\in \lbrack 0,R\rbrack $ obeys the equation
\begin{equation}\label{4}
\sqrt{1-\frac{2m}{r}+\frac{Q^{2}}{r^{2}}}\frac{1}{r}\frac{d}{dr}\left[ 
\sqrt{1-\frac{2m}{r}+\frac{Q^{2}}{r^{2}}}\frac{1}{r}\frac{d\zeta }{dr}%
\right] =\frac{\zeta }{r}\left[ \frac{d}{dr}\frac{m}{r^{3}}+\frac{Q^{2}}{%
r^{5}}\right].
\end{equation}

For $Q=0$ we obtain the equation considered in \cite{St84}.
Since the density $\rho $ does not increase with increasing $r$, the mean
density of the matter $<\rho >=\frac{3m}{4\pi r^{3}}$ inside radius $r$ does
not increase either. Therefore we assume that inside a compact general
relativistic object the condition $\frac{d}{dr}\frac{m}{r^{3}}<0$ holds
independently of the equation of state of dense matter. By defining a new
function
\begin{equation}\label{5}
\eta (r)=\int_{0}^{r}\frac{r^{\prime }}{\sqrt{1-\frac{2m\left( r^{\prime
}\right) }{r^{\prime }}+\frac{Q^{2}\left( r^{\prime }\right) }{r^{\prime 2}}}%
}\left[ \int_{0}^{r^{\prime }}\frac{Q^{2}\left( r^{\prime \prime }\right)
\zeta \left( r^{\prime \prime }\right) }{r^{\prime \prime 5}\sqrt{1-\frac{%
2m\left( r^{\prime \prime }\right) }{r^{\prime \prime }}+\frac{Q^{2}\left( r^{\prime
\prime }\right) }{r^{\prime \prime 2}}}}dr^{\prime \prime }\right]
dr^{\prime },
\end{equation}
denoting $\Psi=\zeta -\eta $, and introducing a new independent variable $\xi =\int_{0}^{r}r^{\prime
}\left( 1-\frac{2m(r^{\prime })}{r^{\prime }}+\frac{Q^{2}\left( r^{\prime
}\right) }{r^{\prime 2}}\right) ^{-\frac{1}{2}}dr^{\prime }$  \cite{St84}, from Eq.(\ref{5}) we obtain the basic result that inside all stable
stellar type charged general relativistic matter distributions the condition 
$\frac{d^{2}\Psi }{d\xi ^{2}}<0$ must hold for all $r\in \left[ 0,R\right] $%
. Using the mean value theorem we conclude $\frac{d\Psi }{d\xi }\leq \frac{%
\Psi \left( \xi \right) -\Psi (0)}{\xi }$, or, taking into account that $%
\Psi (0)>0$ it follows that,
\begin{equation}\label{6}
\Psi ^{-1}\frac{d\Psi }{d\xi }\leq \frac{1}{\xi }.
\end{equation}

In the initial variables the inequality (\ref{6}) takes the form 
\begin{eqnarray}\label{7}
&\frac{1}{r}\left( 1-\frac{2m(r)}{r}+\frac{Q^{2}(r)}{r^{2}}\right) ^{\frac{1%
}{2}}\left[ \frac{1}{2}\frac{d\nu }{dr}e^{\frac{\nu (r)}{2}}-\frac{r}{\sqrt{%
1-\frac{2m}{r}+\frac{Q^{2}}{r^{2}}}}\int_{0}^{r}\frac{Q^{2}\left( r^{\prime
}\right) e^{\frac{\nu \left( r^{\prime }\right) }{2}}}{r^{\prime 5}\sqrt{1-%
\frac{2m\left( r^{\prime }\right) }{r^{\prime }}+\frac{Q^{2}\left( r^{\prime
}\right) }{r^{\prime 2}}}}dr^{\prime }\right] \leq   \nonumber\\
&\frac{e^{\frac{\nu (r)}{2}}-\int_{0}^{r}r^{\prime }\left( 1-\frac{2m\left(
r^{\prime }\right) }{r^{\prime }}+\frac{Q^{2}\left( r^{\prime }\right) }{%
r^{\prime 2}}\right) ^{-\frac{1}{2}}\left[ \int_{0}^{r^{\prime }}\left( 1-%
\frac{2m\left( r^{\prime \prime }\right) }{r^{\prime \prime }}+\frac{%
Q^{2}\left( r^{\prime \prime }\right) }{r^{\prime \prime 2}}\right) ^{-\frac{%
1}{2}}\frac{Q^{2}\left( r^{\prime \prime }\right) e^{\frac{\nu \left(
r^{\prime \prime }\right) }{2}}}{r^{\prime \prime 5}}dr^{\prime \prime
}\right] dr^{\prime }}{\int_{0}^{r}r^{\prime }\left( 1-\frac{2m(r^{\prime })%
}{r^{\prime }}+\frac{Q^{2}\left( r^{\prime }\right) }{r^{\prime 2}}\right)
^{-\frac{1}{2}}dr^{\prime }}.\nonumber\\
\end{eqnarray}

In the following we denote $\alpha (r)=1-\frac{Q^{2}(r)}{2m(r)r}$. For
stable stellar type compact objects $\frac{m}{r^{3}}$ does not increase
outwards. We suppose that for all $r^{\prime }\leq r$ we have $\frac{\alpha
\left( r^{\prime }\right) m(r^{\prime })}{r^{\prime }}\geq \frac{\alpha
\left( r\right) m(r)}{r}\left( \frac{r^{\prime }}{r}\right) ^{2}$ or,
equivalently, $\frac{2m\left( r^{\prime }\right) }{r^{\prime }}-\frac{2m(r)}{%
r}\left( \frac{r^{\prime }}{r}\right) ^{2}\geq \frac{Q^{2}(r^{\prime })}{%
r^{\prime 2}}-\frac{Q^{2}(r)}{r^{2}}\left( \frac{r^{\prime }}{r}\right) ^{2}$%
%.

We assume that inside the compact stellar object the charge $Q(r)$ satisfies
the general condition 
\begin{equation}\label{8}
\frac{Q(r^{\prime \prime })e^{\frac{\nu \left( r^{\prime \prime }\right) }{2}%
}}{r^{\prime \prime 5}}\geq \frac{Q(r^{\prime })e^{\frac{\nu \left(
r^{\prime }\right) }{2}}}{r^{\prime 5}}\geq \frac{Q(r)e^{\frac{\nu (r)}{2}}}{%
r^{5}},r^{\prime \prime }\leq r^{\prime }\leq r.
\end{equation}

Therefore we can evaluate the terms in equation (\ref{7}) as follows. For
the term in the denominator of the RHS of Eq.(\ref{7}) we obtain: 
\begin{equation}\label{9}
\left[ \int_{0}^{r}r^{\prime }\left( 1-\frac{2m\left( r^{\prime }\right) }{%
r^{\prime }}+\frac{Q^{2}\left( r^{\prime }\right) }{r^{\prime 2}}\right) ^{-%
\frac{1}{2}}dr^{\prime }\right] ^{-1}\leq \frac{2\alpha (r)m(r)}{r^{3}}%
\left[ 1-\left( 1-\frac{2\alpha (r)m(r)}{r}\right) ^{\frac{1}{2}}\right]
^{-1}.
\end{equation}

For the second term in the bracket of the LHS of Eq.(\ref{7}) we have 
\begin{eqnarray}\label{10}
&\int_{0}^{r}\left( 1-\frac{2m(r^{\prime })}{r^{\prime }}+\frac{Q^{2}\left(
r^{\prime }\right) }{r^{\prime 2}}\right) ^{-\frac{1}{2}}\frac{Q^{2}\left(
r^{\prime }\right) e^{\frac{\nu \left( r^{\prime }\right) }{2}}}{r^{\prime 5}%
}dr^{\prime }\geq   \nonumber \\
&\frac{Q^{2}(r)e^{\frac{\nu (r)}{2}}}{r^{5}}\int_{0}^{r}\left( 1-\frac{2m(r)%
}{r^{3}}r^{\prime 2}+\frac{Q^{2}(r)}{r^{6}}r^{\prime 2}\right) ^{-\frac{1}{2}%
}dr^{\prime }=  \nonumber\\
&\frac{Q^{2}(r)e^{\frac{\nu (r)}{2}}}{r^{5}}\left( \frac{2\alpha (r)m(r)}{%
r^{3}}\right) ^{-\frac{1}{2}}\arcsin \left( \sqrt{\frac{2\alpha (r)m(r)}{r}}%
\right).\nonumber \\
\end{eqnarray}

The second term in the bracket of the RHS of Eq. (\ref{7}) can be evaluated
as 
\begin{eqnarray}\label{11}
&\int_{0}^{r}r^{\prime }\left( 1-\frac{2m\left( r^{\prime }\right) }{%
r^{\prime }}+\frac{Q^{2}\left( r^{\prime }\right) }{r^{\prime 2}}\right) ^{-%
\frac{1}{2}}\left[ \int_{0}^{r^{\prime }}\left( 1-\frac{2m\left( r^{\prime
\prime }\right) }{r^{\prime \prime }}+\frac{Q^{2}\left( r^{\prime \prime
}\right) }{r^{\prime \prime 2}}\right) ^{-\frac{1}{2}}\frac{Q^{2}\left(
r^{\prime \prime }\right) e^{\frac{\nu \left( r^{\prime \prime }\right) }{2}}%
}{r^{\prime \prime 5}}dr^{\prime \prime }\right] dr^{\prime }\geq \nonumber \\
&\int_{0}^{r}r^{\prime }\left( 1-\frac{2m(r^{\prime })}{r^{\prime }}+\frac{%
Q^{2}\left( r^{\prime }\right) }{r^{\prime 2}}\right) ^{-\frac{1}{2}}\frac{%
Q^{2}\left( r^{\prime }\right) e^{\frac{\nu \left( r^{\prime }\right) }{2}}}{%
r^{\prime 4}}\left( \frac{2\alpha (r^{\prime })m(r^{\prime })}{r^{\prime }}%
\right) ^{-\frac{1}{2}}\arcsin \left( \sqrt{\frac{2\alpha (r^{\prime
})m(r^{\prime })}{r^{\prime }}}\right) dr^{\prime }\geq   \nonumber \\
&\frac{Q^{2}(r)e^{\frac{\nu (r)}{2}}}{r^{5}}\int_{0}^{r}r^{\prime }\left( 1-%
\frac{2\alpha (r)m(r)}{r}r^{\prime 2}\right) ^{-\frac{1}{2}}\left( \frac{%
2\alpha (r)m(r)}{r^{3}}r^{\prime 2}\right) ^{-\frac{1}{2}}\arcsin \left( 
\sqrt{\frac{2\alpha (r)m(r)}{r^{3}}}r^{\prime }\right) dr^{\prime }= \nonumber \\
&\frac{Q^{2}(r)e^{\frac{\nu (r)}{2}}}{r^{5}}r^{2}\left( \frac{2\alpha
(r)m(r)}{r}\right) ^{-\frac{3}{2}}\left[ \sqrt{\frac{2(r)m(r)}{r}}-\sqrt{1-%
\frac{2\alpha (r)m(r)}{r}}\arcsin \left( \sqrt{\frac{2\alpha (r)m(r)}{r}}%
\right) \right].\nonumber \\
\end{eqnarray}

In order to obtain the inequality (\ref{11}) we have also used the property
of monotonic increase in the interval $x\in \left[ 0,1\right] $ of the
function $\frac{\arcsin x}{x}$ .

Using Eqs.(\ref{9})-(\ref{11}), Eq.(\ref{7}) becomes: 
\begin{equation}\label{12}
\left[ 1-\left( 1-\frac{2\alpha (r)m(r)}{r}\right) ^{\frac{1}{2}}\right] 
\frac{m(r)+4\pi r^{3}p-\frac{Q^{2}}{r}}{r^{3}\sqrt{1-\frac{2\alpha (r)m(r)}{r%
}}} \leq \frac{2m(r)}{r^{3}}+\frac{Q^{2}}{r^{4}}\left[ \frac{\arcsin \left( 
\sqrt{\frac{2\alpha (r)m(r)}{r}}\right) }{\sqrt{\frac{2\alpha (r)m(r)}{r}}}%
-1\right].
\end{equation}

Eq. (\ref{12}) is valid for all $r$ inside the star. Consider first the
neutral case $Q=0$. By evaluating (\ref{12}) for $r=R$ we obtain $\frac{1}{%
\sqrt{1-\frac{2M}{R}}}\leq 2\left[ 1-\left( 1-\frac{2M}{R}\right) ^{\frac{1}{%
2}}\right] ^{-1},$ leading to the well-known result $\frac{2M}{R}\leq \frac{8%
}{9}$ \cite{St84}.

Next consider the case $Q\neq 0$. We denote 
\begin{equation}\label{13}
f\left( M,R,Q\right) =\frac{Q^{2}(R)}{R^{4}}\left( \frac{\alpha (R)M}{R^{3}}%
\right) ^{-1}\sqrt{1-\frac{2\alpha (R)M}{R}}\left[ \frac{\arcsin \left( 
\sqrt{\frac{2\alpha (R)M}{R}}\right) }{\sqrt{\frac{2\alpha (R)M}{R}}}%
-1\right].
\end{equation}

The function $f\left( M,R,Q\right) \geq 0,\forall M,R,Q$. Then (\ref{12})
leads to the following restriction on the mass-radius ratio for compact
charged general relativistic objects: 
\begin{equation}\label{14}
\frac{2M}{R}\leq \frac{8}{9}+\frac{2f\left( M,R\right) }{9}-\frac{%
f^{2}\left( M,R\right) }{9}.
\end{equation}

The variation of the maximum mass-radius ratio $u=\frac{M}{R}$ for a charged compact object
as a function of the charge-mass ratio $q=\frac{Q}{M}$ is represented in Fig.1.

\begin{figure}[hb]
\epsfxsize=10cm
\centerline{\epsffile{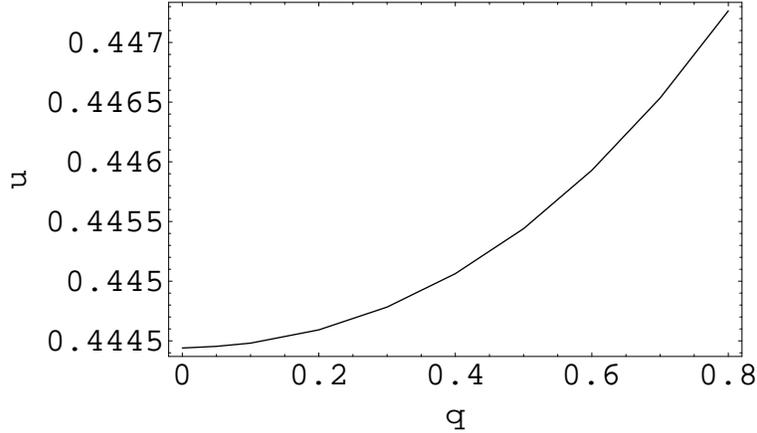}}
\caption{Variation of the maximum value $u$ of the mass-radius ratio $M/R$ for
charged compact general-relativistic objects as a function of the total charge-mass ratio
$q=Q/M$.}
\label{FIG1}
\end{figure}
 
Due to the presence of the charge the maximum mass-radius ratio is only slightly modified
as compared to the non-charged case. In the uncharged case the bound $2\frac{M}{R}<\frac{8}{9}$
is very close to the limit $2\frac{M}{R}<1$ arising from black hole considerations. But
for a charged compact general relativistic object the bound $\frac{M}{R}<1$
obtained from horizon considerations is much larger than the limit
following from Eq.(\ref{14}).

In order to find a general restriction for the total charge $Q$ a compact
stable object can acquire we shall consider the behavior of the Ricci
invariants $r_{0}=R_{i}^{i}=R$, $r_{1}=R_{ij}R^{ij}$ and $%
r_{2}=R_{ijkl}R^{ijkl}$. If the general static line element is regular,
satisfying the conditions $e^{\nu (0)}=const.\neq 0$ and $e^{\lambda (0)}=1$
, then the Ricci invariants are also non-singular functions throughout the
star. In particular for a regular space-time the invariants are
non-vanishing at the origin $r=0$. For the invariant $r_{2}$ we find 
\begin{eqnarray}\label{15}
&r_{2}=\left[ 8\pi \left( \rho +p\right) -\frac{4m}{r^{3}}+\frac{6Q^{2}}{%
r^{4}}\right] ^{2}+2\left( 8\pi p+\frac{2m}{r^{3}}-\frac{2Q^{2}}{r^{4}}%
\right) ^{2}+  \nonumber \\
&2\left( 8\pi \rho -\frac{2m}{r^{3}}+\frac{2Q^{2}}{r^{4}}\right)
^{2}+4\left( \frac{2m}{r^{3}}-\frac{Q^{2}}{r^{4}}\right) ^{2}. \nonumber \\
\end{eqnarray}

For a monotonically decreasing interior electric field $\frac{Q^{2}}{8\pi
r^{4}}$, the function $r_{2}$ is regular and motonically decreasing
throughout the star. Therefore it satisfies the condition $r_{2}(R)<r_{2}(0)$%
, leading to the following general constraint on the value of the electric
field at the surface of the compact object: 
\begin{equation}\label{16}
6\frac{M^{2}}{R^{6}}<\frac{12M}{R^{3}}\frac{Q^{2}}{R^{4}}-7\left( \frac{Q^{2}%
}{R^{4}}\right) ^{2}+4\pi ^{2}\left( 6\rho _{c}^{2}+4\rho
_{c}p_{c}+6p_{c}^{2}\right),
\end{equation}
where we assumed that at the surface of the star the matter density
vanishes, $\rho (R)=0$.

Another condition on $Q(R)$ can be obtained from the study of the scalar 
\begin{equation}\label{17}
r_{1}=\left( 8\pi \rho +\frac{Q^{2}}{r^{4}}\right) ^{2}+3\left( 8\pi p-\frac{%
Q^{2}}{r^{4}}\right) ^{2}+\frac{64\pi pQ^{2}}{r^{4}}.
\end{equation}

Under the same assumptions of regularity and monotonicity for the function $%
r_{1}$ and considering that the surface density is vanishing we obtain for
the surface value of the monotonically decreasing electric field the upper
bound 
\begin{equation}\label{18}
\frac{Q^{2}}{R^{4}}<4\pi \rho _{c}\sqrt{1+3\left( \frac{p_{c}}{\rho _{c}}%
\right) ^{2}}.  
\end{equation}

The invariant $r_{0}$ leads to the trace condition $\rho _{c}>3p_{c}$ of the energy-momentum tensor that holds at the center of 
the fluid spheres.

\section{Discussions and final remarks}

The existence of a limiting value of the mass-radius ratio leads to limiting
values for other physical quantities of observational interest. One of these
quantities is the surface red shift $z$ of the compact object, defined
according to $z=\left( 1-\frac{2M}{R}\right) ^{-\frac{1}{2}}-1$. For an electrically neutral star Eq.(\ref{7}) leads to the well-known constraint $z\leq 2$.
For the charged star the surface red shift must obey the more general
restriction
\begin{equation}\label{20}
z\leq 2+\frac{Q^{2}}{R^{4}}\left( \frac{\alpha (R)M}{R^{3}}\right)
^{-1}\left[ \frac{\arcsin \left( \sqrt{\frac{2\alpha (R)M}{R}}\right) }{%
\sqrt{\frac{2\alpha (R)M}{R}}}-1\right].
\end{equation}

The variation as a function of the charge-mass ratio of the maximum red shift
for charged compact objects for a given mass-radius ratio is presented in Fig.2.

\begin{figure}[hb]
\epsfxsize=10cm
\centerline{\epsffile{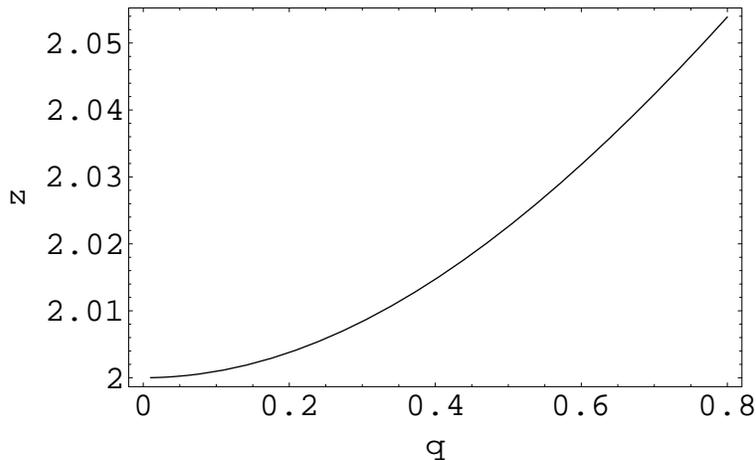}}
\caption{Variation of the maximum red shift $z$ of a compact general
relativistic object as a function of the total charge-mass ratio
$q=Q/M$ for a value of the mass-radius ratio of $u=0.4$.}
\label{FIG2}
\end{figure}

Therefore higher surface red shifts than 2 could be observational
criteria indicating the presence of electrically charged ultra compact
matter distributions.

As another application of obtained limiting mass -radius ratios for charged
stars we shall derive an explicit limit for the total energy of compact
general relativistic objects. The total energy (including the gravitational
field contribution) inside an equipotential surface $S$ can be defined to be \cite{Ka88}
\begin{equation}\label{21}
E=E_{M}+E_{F}=\frac{1}{8\pi }\xi _{s}\int_{S}\left[ K\right] dS,
\end{equation}
where $\xi ^{i}$ is a Killing field of time translation, $\xi _{s}$ its
value at $S$ and $\left[ K\right] $ is the jump across the shell of the
trace of the extrinsic curvature of $S$, considered as embedded in the
2-space $t=const.$.$E_{M}=\int_{S}T_{i}^{k}\xi ^{i}\sqrt{-g}dS_{k}$ and $%
E_{F}$ are the energy of the matter and of the gravitational field,
respectively. This definition is manifestly coordinate invariant. In the
case of static spherically symmetric matter distribution we obtain for the
total energy (also including the gravitational contribution) the exact
expression $E=-re^{\frac{\nu -\lambda }{2}}$ \cite{Ka88}.
Hence the total energy (including the gravitational contribution) of a
charged compact general relativistic object is $E=-R\left[ 1-\frac{2M}{R}+\frac{Q^{2}(R)}{R^{2}}\right] $

For a neutral matter distribution $Q=0$ and for the total energy of the star
we find the upper limit $E\leq -\frac{R}{9}$. In the charged case we obtain
\begin{equation}\label{22}
E\leq -\frac{R}{9}+\frac{2f}{9}R-\frac{f^{2}}{9}R,
\end{equation}
with the function $f$ defined in Eq. (\ref{13}).

In the present Letter we have considered the mass-radius ratio limit for
charged stable compact general relativistic objects. Also in this case it is
possible to obtain explicit inequalities involving $\frac{2M}{R}$ as an
explicit function of the charge $Q$. The surface red shift and the total
energy (including the gravitational one) are modified due to the
presence of a strong electric field inside the compact object. The
mass-radius ratio depends on the value of the total charge
of the star, with the increases in mass, red shift or total energy proportional
to the charge parameter.


\begin{references}

\bibitem{Ch99} Chandrasekhar S., {\it The Mathematical Theory of Black Holes}, (1992) Oxford, Oxford University Press.

\bibitem{Sh71} Shvartsman V.F., {\it Soviet Physics-JETP}, {\bf 33} 1971 475.

\bibitem{Tu97} Turolla R., Zane S., Treves A. and  Illarionov A., {\it Astrophys. J.}, {\bf 482} (1997) 377;
 Michel F.C., {\it Astrophys. Space Science}, {\bf 15} (1972) 153;
 Anile A.M. and Treves A., {\it Astrophys. Space Science}, {\bf 19} (1972) 411;
 Maraschi I., Reina C. and  Treves A., {\it Astron. Astrophys.}, {\bf 35} (1974) 389;
 Meszaros P., {\it Astron. Astrophys.}, {\bf 44} (1975) 59;
 Maraschi I., Reina C. and Treves A., {\it Astron. Astrophys.}, {\bf 66} (1978) 99;
 Treves A. and Turolla R., preprint astro-ph/9812383 (1998).

\bibitem{Pr98}  Preparata G.,  Ruffini R. and  Xue S.-S., preprint astro-ph/9810182 (1998); Ruffini R., preprint astro-ph/9811232 (1998);
 Ruffini R., Salmonson J.D., Wilson J.R.and  Xue S.-S., preprint astro-ph/0004257 (2000); Ruffini R., preprint astro-ph/0001425 (2000).

\bibitem{Wi84} Witten E., {\it Phys. Rev. D}, {\bf 30} (1984) 272.

\bibitem{Kh95} Khadrikar S.B., Mishra A. and Mishra H., {\it Mod. Phys. Lett. A}, {\bf 10} (1995) 2651.

\bibitem{Bu59}  Buchdahl H.A., {\it Phys. Rev.}, {\bf 116} (1959) 1027.

\bibitem{St84} Straumann N., {\it General Relativity and Relativistic Astrophysics} (1984), Springer Verlag, Berlin. 
 
\bibitem{Be71} Beckenstein J.D., {\it Phys. Rev. D}, {\bf 4} (1971) 2185.

\bibitem{Ka88} Katz J., Lynden-Bell D. and Israel W., {\it Class. Quantum. Grav.}, {\bf 5} (1988) 971; Gron O. and  Johannesen S., {\it Astrophys. Space Science}, {\bf 19} 1992 411.

\end{references}
\end{document}